\documentclass[prl,letterpaper,twocolumn,tightenlines,superscriptaddress,showpacs,preprintnumbers,nofootinbib,floatfix,psfig]{revtex4-1}

\usepackage{amsmath,amssymb,amsfonts,graphicx,pifont,dcolumn}
\usepackage{epstopdf}

\RequirePackage{slashed}

\newcommand{\bit}{\begin{itemize}}
\newcommand{\eit}{\end{itemize}}

\def\be{\begin{equation}}
\def\ee{\end{equation}}
\def\bea{\begin{eqnarray}}
\def\eea{\end{eqnarray}}

\begin{document}

%\preprint{UNITN 11-00}

\author{Marco Traini}

\affiliation{Dipartimento di Fisica, Universit\`a degli Studi di Trento, I-38100 Povo (Trento), Italy\\
and INFN, Gruppo Collegato di Trento, sezione di Padova\\(e-mail:traini@science.unitn.it)}

\title{Charge symmetry violation: \\ a NNLO study of partonic observables}

\begin{abstract}
Charge and isospin symmetry violations to valence and sea distribution functions in the nucleon are evaluated (at low resolution scale) by means of a meson cloud model and light-cone quark wave functions. Their perturbative evolution are implemented at Next-to-next-to-leading order (NNLO) using an original evolution code in order to include the perturbatively generated strange - anti-strange asymmetry typical of the three loop evolution expansion. Charge symmetry violating QED effects are also added and the distributions, evolved at the experimental scale, are compared with available information. The role of non-perturbative effects is emphasized in the interpretation of the, so called,  NuTeV anomaly and new experiments at very-high energy.
\end{abstract}

\maketitle

%\begin{center}
%\today
%\end{center}

\noindent {\bf 1. introduction}

%\vspace{0.1truecm}

At low energy the nuclear forces are commonly described within a symmetry property strictly valid when the electromagnetic interactions are turned off: charge symmetry. The small differences in the masses of the neutron and the proton ($(M_n - M_p) / (M_n + M_p) \approx 0.1\%$) are the documentation of the validity of such a property.

At the partonic level charge symmetry implies [e.g. ref.\cite{CSV2010}] that the distributions
\begin{eqnarray}
\delta u(x,Q^2) & = & u^p(x,Q^2) - d^n(x,Q^2) = \nonumber \\
& = & \delta u_V(x,Q^2) + \delta \bar u(x,Q^2)\,,\nonumber \\
\delta d(x,Q^2) & = & d^p(x,Q^2) - u^n(x,Q^2) = \nonumber \\
& = & \delta d_V(x,Q^2) + \delta \bar d(x,Q^2)\,; \label{CSpartonicV} \\
{\rm with} \;\;\;\;\;\;\;\; && \nonumber \\
\delta \bar u(x,Q^2) & = & \bar u^p(x,Q^2) - \bar d^n(x,Q^2)\,,\nonumber \\
\delta \bar d(x,Q^2) & = & \bar d^p(x,Q^2) - \bar u^n(x,Q^2)\,; \label{CSpartonicSea} 
\end{eqnarray}
vanish identically. However relations (\ref{CSpartonicV}), (\ref{CSpartonicSea}) are broken by the $u - d$ mass difference as demonstrated (for the valence sector of eqs.(\ref{CSpartonicV})) by Sather \cite{Sather1992} and Rodionov {\sl et al} \cite{Rodionovetal1994} by means of a MIT bag model.
Those authors predict similar behavior and magnitude for $\delta d_V$ and $\delta u_V$ (as large as few percent  at intermediate $x$-values), but with opposite sign. Rather recently Martin {\sl et al} (MRST) \cite{MRST2004} proposed a global phenomenological analysis of PDFs which include charge symmetry violation (CSV) contribution: their best fit (with large uncertainties) show effects similar to the predictions by Sather and Rodionov.

A more recent and theoretical estimate of the second moments of the CSV valence contributions have been performed by Horsley {\sl et al} \cite{Horsleyetal2011} within a lattice QCD approach, obtaining\\
%\begin{eqnarray}
$\langle x \delta u_V(x) \rangle  = \int dx\, x\, \delta u_V(x,Q^2=4\,{\rm GeV^2}) = -0.0023(6)$  \\
$\langle x \delta d_V(x) \rangle = \int dx\, x\, \delta d_V(x,Q^2=4\,{\rm GeV^2}) = +0.0020(3)\,,$
%\end{eqnarray}
also in agreement with the best fit of MRST \cite{MRST2004} ($\langle x \delta u_V(x) \rangle = - \langle x \delta d_V(x) \rangle  = -0.002^{+0.009}_{-0.006}$ [90\% CL]) and MIT bag model predictions.

However at the same level of accuracy one should consider, in addition to mass difference effects of the order $(m_d - m_u) / \Lambda_{\rm QCD}$, electromagnetic (QED) contributions of the order $\alpha_{EM}$ as already proposed by Sather \cite{Sather1992};  both are important in fixing CSV effects, in particular to constrain parton distribution functions and hence the accuracy with which (for example) cross section at the LHC can be predicted, or to guide the search for physics beyond Standard Model. Inclusion of QED splitting contribution have been considered within the MRST phenomenological fit \cite{MRST2004}, and by Gl\"uck, Jimenez-Delgado and Reya \cite{GJDR2005}. Their approach is based on the inclusion of photon radiation effects in the QCD evolution equations: the explicit coupling of quarks to photon being analogous to the coupling of quarks to gluon. The approach allows the explicit evaluation of the CSV in the valence sector (eqs. (\ref{CSpartonicV})), and in the sea sector (eqs.(\ref{CSpartonicSea})) enlarging to new CSV effects at the partonic level. 

The interest in small partonic CSV effects is not only speculative, all the corrections discussed above are, in fact, relevant for the interpretation of the so called NuTeV anomaly, i.e., the reported (on shell) measurement of the Weinberg angle \cite{NuTeVdata} ($\sin^2 \theta_W = 0.2277 \pm 0.0013\,(stat)  \pm 0.0009\,(syst)$) which is (approximatly) three standard deviations {\it above} the world fit to other electroweak processes ($0.2227 \pm 0.0004$) (cfr. \cite{CSV2010} and references therein). Although corrections beyond the standard model have been discussed \cite{B-standard-M}, they appear to be rather speculative \cite{CSV2010}, while a certain number of more standard corrections must be considered (e.g. refs.\cite{CSV2010,Bentz-etal2010}). 

Integrating  all the CSV effects coming from mass differences and electromagnetic radiation in a new and common framework, is a first aim of the present work offering a theoretical approach which can show, within a unified picture,  the relative importance of different essential CSV contributions. 
A second aim of the present paper is offered by study of high energy experiments to measure CSV effects.
In particular it has been proposed \cite{Hobbs_etal2011} to measure CSV by comparing neutrino or antineutrino production through charged-current reactions induced by electrons or positrons at a possible electron collider at the LHC (LHeC \cite{http_uk}): predictions within the present approach will be presented.
 
\vspace{0.1truecm}

\noindent{\bf 2. NuTeV corrections}

%\vspace{0.1truecm}

The NuTeV experiment used a steel target to extract the ratios of the neutral current (NC) to charged current (CC) total cross sections for neutrinos ($R^\nu$) and anti-neutrinos ($R^{\bar \nu}$). The extraction of $\sin^2 \theta_W$ is the result of a Monte Carlo simulation of the experiment. The Paschos-Wolfenstwein (PW) relation \cite{PW1973} (valid for an isoscalar target) establishes the connection with the weak mixing angle. One has for the PW ratio $R_{PW}$ ($s^2_W \equiv \sin^2 \theta_W$):
\begin{eqnarray}
&& R_{PW}  =  {\langle \sigma_{NC}^{\nu \, A}\rangle  - \langle \sigma_{NC}^{\bar \nu \, A}\rangle  \over \langle  \sigma_{CC}^{\nu\,A} \rangle -  \langle \sigma_{CC}^{\bar \nu\,A}\rangle} \equiv {R^\nu - r \, R^{\bar \nu } \over 1- r} =  \label{PWratio}\\
& = & {({1 \over 6} - {4 \over 9}s^2_W) \,\langle x_A \, u_A^- \rangle + ({1 \over 6} - {2 \over 9}s^2_W) \,\langle x_A\, d_A^-  + x_A \, s_A^- \rangle \over \langle x_A \, d_A^- + x_A \, s_A^- \rangle - {1 \over 3} \langle x_A \, u_A^-\rangle}\,, \nonumber \\ \label{partonicA}
\end{eqnarray}
and the cross sections have been integrated over the Bjorken scaling variable and energy transfer.  A represents the nuclear target and $r = {\langle \sigma_{CC}^{\bar \nu\,A}\rangle / \langle \sigma_{CC}^{\nu\,A}\rangle }$. Eq.(\ref{PWratio})
expresses the Paschos-Wolfenstein observation, and eq.(\ref{partonicA}) shows the explicit dependence of the cross sections on the quark distributions (ignoring the heavy quark flavours and ${\cal O}(\alpha_s)$ corrections \cite{Bentz-etal2010}):
$x_A$ is the Bjorken scaling variable for the nucleus A and $q^-_A \equiv q_A - \bar q_A$ are the non-singlet quark distributions of the target. Expanding eq.(\ref{partonicA}) to include CSV and strange quark effects one obtains the leading PW corrections:
\begin{eqnarray}
&& R_{PW}  =  {1 \over 2} - \sin^2 \theta_W + \left(1 - {7 \over 3}s^2_W \right)\, {1 \over \langle x\,u_V\rangle + \langle x\,d_V \rangle} \times \nonumber \\
& \times & \left\{ - {N - Z \over A} \left[ \langle x\,u_V \rangle - \langle x\,d_V \rangle \right] + {N \over A}\left[
\langle x \, \delta u_V \rangle + \right.\right. \nonumber \\
& \phantom{\times}&Ê \left. \left. \; - \langle x \, \delta d_V \rangle \right] - \langle x \, (s - \bar s) \rangle \phantom{{}N \over A} \right\} = \nonumber \\
& = & 
{1 \over 2} - \sin^2 \theta_W + \Delta R^{N,Z}_{PW } + \Delta R^{CSV}_{PW }+ \Delta R^{\rm strange}_{PW }\,;\label{PWexpansion}
\end{eqnarray}
where 
\begin{eqnarray}
\langle x_A \, q_A \rangle & = & N \, \langle x \, q ^n \rangle + Z \, \langle x \, q^p \rangle
\nonumber \\
\langle x_A \, \bar q_A \rangle & = & N \, \langle x \, \bar q^n \rangle + Z \, \langle x \, \bar q^p \rangle \,, \label{incoherent}\\
{\rm and} \;\;\;\;\; &&\nonumber \\ 
\langle x_A \, s^-_A \rangle & = & (N+Z) \langle x\, s^- \rangle\,,\label{strange_np}
\end{eqnarray}
have been assumed: i.e. an identical strange asymmetry for neutron and proton (\ref{strange_np}) and  incoherent scattering on the single parton in the nucleus (\ref{incoherent}). We will come back on the assumption (\ref{incoherent}) later on the paper.
Eq.(\ref{PWexpansion}) shows relevant corrections to the PW result ($\left. R_{PW}\right|_{N=Z} = {1 \over 2} - \sin^2 \theta_W$) valid when quark mass differences, electroweak corrections (CSV), strange quark effects are neglected, and the target is purely isoscalar ($N=Z$). In the following the contributions $ \Delta R^{N,Z}_{PW }, \Delta R^{CSV}_{PW }, \Delta R^{\rm strange}_{PW }$ will be explicitly evaluated and discussed, and it is worthy to note that the CSV term is also renormalized by the non isoscalar nature of the target 
\begin{equation}
{1 \over 2}\,{\langle x \, \delta u_V \rangle - \langle x \, \delta d_V \rangle  \over \langle x \, u_V \rangle + \langle x \,  d_V \rangle} \to {N \over A} \,{\langle x \, \delta u_V \rangle - \langle x \, \delta d_V \rangle  \over \langle x \, u_V \rangle + \langle x \,  d_V \rangle} \,,\label{N/A}
\end{equation}
 a correction due to neutron excess overlooked till now (e.g cfr. eq.(58) in \cite{CSV2010}, eqs.(8) in \cite{GJDR2005}, eq.(25) in \cite{SongZhangMa2010}).

\vspace{0.1truecm}

\noindent{\bf 3. the model}

%\vspace{0.1truecm}

Following one the main aim of the present work, the partonic contributions to eq.(\ref{PWexpansion}) will be evaluated within a quark-parton model which can give a self-consistent evaluation of all the ingredients and corrections. The model assumes that the twist-two component of the parton distributions can be calculated within a relativistic light-front constituent quark model whose Hamiltonian is written by means of an hypercentral potential between quarks 
\cite{FaccioliTrainiVento1999}. 

According to that approach the parton distribution takes the simple form: 
\begin{eqnarray}
\label{LFq-val}
q(x,\mu_0^2) & = & \sum_{j=1}^3\delta_{\tau_j\tau_q} \int 
\prod_{i=1}^3 d\vec{k}_i \, \delta\left(\sum_{i=1}^3 \vec{k}_i\right)
\,\delta\left(x-\frac{k^+_j}{M_0}\right) \times \nonumber \\
& & \phantom{\sum_{j=1}^3\delta_{\tau_j\tau_q \int \prod_{i=1}^3 d\vec{k}_i \, } }Ê\times \vert\Psi_\lambda^{[c]}(\{\vec{k}_i;\lambda_i,\tau_i\})\vert^2\,,\label{LFmodel}
\end{eqnarray}
where $k_j^+=(k^0_j+k_j^3)/\sqrt{2}$ is
the quark light-cone momentum, and $M_0=\sum_i\, \sqrt{{\vec k}^2_i+m^2_i}$ is 
the free mass for the three-quark system. 
$\Psi_\lambda^{[c]}(\{\vec{k}_i;\lambda_i,\tau_i\})$ is the canonical wave
function of the nucleon in the instant form obtained by solving an eigenvalue equation for
the mass operator:
$M = \sum_{i=1}^3 \sqrt{\vec{k}_i^2 + m_i^2} -\frac{\tau}{y} +\kappa_l \,y,$, with
$\sum_i\vec{k}_i = 0$. $m_i$ is the constituent quark masses,
$y=\sqrt{\vec{\rho}^2 + \vec{\lambda}^2}$ is the radius of the
hypersphere in six dimensions and $\vec{\rho}$ and $\vec{\lambda}$ are 
Jacobi coordinates (refs.~\cite{FaccioliTrainiVento1999,GE95}).

The distribution (\ref{LFq-val}) automatically fulfills the support condition and satisfies the (particle)
baryon number and momentum sum rules at the hadronic scale $\mu_0^2$  
where the valence contribution dominates the twist-two response \cite{PasquiniTrainiBoffi2005}.
The core quark model, just described, is surrounded by a cloud of mesons to incorporate $q \bar q$ non-perturbative components. The physical nucleon state is expanded 
(in the infinite momentum frame (IMF) and in the one-meson approximation) in a series involving bare nucleons 
and two-particle, meson-baryon states (e.g. \cite{MCMold}).

In DIS the virtual photon can hit either the bare proton
$p$ or one of the constituents of the higher Fock states. In the IMF, where the
constituents of the target can be regarded as free during the interaction time,
the contribution of the higher Fock states to the quark distribution of the
physical proton, can be written as the convolution \\
%\begin{eqnarray}
 $\Delta q_{p}(x)  =  \sum_{MB} \left [
  \int_x^1 \frac{dy}{y}\,f_{MB/p}(y)\, q_M \left(\frac{x}{y}\right) +\right.$

\noindent $\left.+ \int_x^1 \frac{dy}{y}\, f_{BM/p}(y)\, q_B \left(\frac{x}{y}\right)
 \right ]$ , 
 %\label{eq:splitting}
%\end{eqnarray}
where the splitting functions $f_{MB/p}(y)$ and $f_{BM/p}(y)$ are related to the
probability amplitudes for the proton to
fluctuate into a virtual baryon-meson system ($BM$) with the baryon and meson
having longitudinal momentum fractions $y$ and $1-y$,
($f_{BM/p}(y) = f_{MB/p}(1-y)$. They are calculated by using time-order perturbation theory
in the IMF.
The quark distributions in a {\em physical} proton are then given by
%\begin{equation}
$q_{p} (x) = Z q_{p}^{\rm bare}(x) + \Delta q_{p}(x) $, 
%\label{eq:partondis}
%\end{equation}
where $q_{p}^{\rm bare}$ are the bare quark distributions and the renormalization
constant. $ Z\equiv 1 - \sum_{MB} \int_0^1  dy\,f_{MB/p}(y)$ is equal
to the probability of finding the bare proton in the physical proton. 
For all the details see ref.\cite{PasquiniTrainiBoffi2005}, where the model
has been used to calculate the non-perturbative components of Generalized proton Parton Distribution.

%===============================================================================
\begin{figure}[tbp]
\centering\includegraphics[width=\columnwidth,clip=true,angle=0]{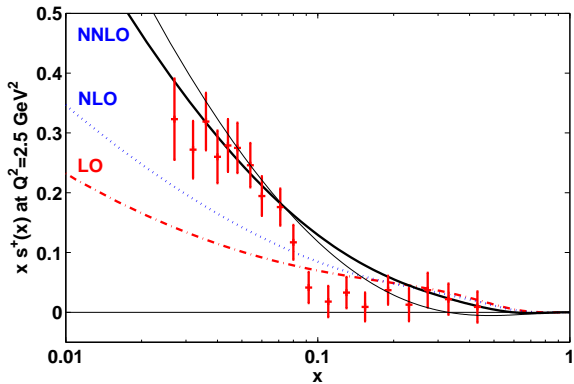}
\caption{$x s^+(x,Q^2) = x\,(s(x,Q^2) + \bar s(x,Q^2))$ at $Q^2 = 2.5\,{\rm GeV^2}$, the scale of the Hermes data which are shown for comparison and adapted from ref.\cite{HERMES2008}. Dot-dashed line is the result of a LO evolution, dotted line the NLO evolution, continuous lines the NNLO evolution for scenario {\bf B} where calculations include non-perturbative strange sea at the static evolution scale. Comparison with results with vanishing strange sea at the starting scale (scenario {\bf A}) is made (at NNLO) by means of the tiny continuous line. Let us recall that the strange asymmetries at NLO an LO would be strictly zero if the strange sea is assumed to vanish at $Q_0^2$.
}
\label{fig:xsplus}
\vspace{-1.0em}
\end{figure}
%===============================================================================%%%%%

%\vspace{0.1truecm}

\noindent{\bf 4. QCD evolution and QED contributions}

%\vspace{0.1truecm}

In the following we will assume that, at the lowest hadronic scale, the bare
nucleon is described by the relativistic quark model wave function formulated
within the light-front dynamics and, as a consequence, only valence partons
will contribute to the partonic content of the bare nucleon. The inclusion of the meson cloud will  be mainly done adding non-strange sea (scenario {\bf A}), and  proton fluctuations in virtual states  of  $\Delta$ and $\pi$  are explicitly taken into account \cite{PasquiniTrainiBoffi2005}. 
An attempt to add strange component to the meson cloud is also investigated.
In that case the strange component will be added following the prescription by Melnitchouck and Malheiro \cite{MeMa1997}  and by Chen, Cao and Signal \cite{ChenCaoSignal2010}; proton fluctuations in virtual states of  $\Lambda$ and $K$ \cite{MeMa1997,ChenCaoSignal2010} will be explicitly considered (scenario {\bf B}).
The distributions calculated within the light-front (meson-cloud) model are used as starting (static) point of a QCD evolution to the experimental  scale. The processes involved require evolution where the strange asymmetry is well identified and its properties well described. Because of such considerations an original Next-to-next-to-Leading Order (NNLO) evolution code will be used \cite{Traini2011} which includes the perturbatively generated strange - antistrange asymmetry typical of the three loop evolution expansion \cite{Catani_etal2004}.

The two scenarios have different valence and sea relative contributions
and, therefore, exhibit different scale; {\bf A}: $Q_0^2 = 0.149\,{\rm Gev^2}$ if only $\bar u$ and $\bar d$ are considered ($\langle x [ 2 \bar u(x,Q_0^2) + 2 \bar d(x,Q_0^2)] \rangle = 0.110 ; \langle x [u_V(x,Q_0^2) + d_V(x,Q_0^2)] \rangle = 0.890$); {\bf B}: $Q_0^2 =  0.161\,{\rm GeV^2}$ if the strange sea is included  ($\langle x [ 2 \bar u(x,Q_0^2) + 2 \bar d(x,Q_0^2)+s(x,Q_0^2) + \bar s (x,Q_0^2)] \rangle = 0.151 ; \langle x [u_V(x,Q_0^2) + d_V(x,Q_0^2)] \rangle = 0.849$). In particular the model distributions show a strange asymmetry $\left. \langle x (s - \bar s)\rangle \right|_{Q_0^2} = - 0.0057$ within the scenario {\bf B} while it vanishes within the model {\bf A}.
Evolving, at NNLO \cite{note_a_s},  to $Q^2 = 4\,{\rm GeV^2}$ one obtains $\left. \langle x (s - \bar s) \rangle \right|^{\bf A}_{Q^2} = - 0.0039$, while if the strange sea is vanishing at the static point one gets $\left. \langle x (s - \bar s)\rangle \right|_{Q^2} = - 0.0015$. A rather {\em large and negative} value which is fully compatible with the experimental results of the Hermes collaboration \cite{HERMES2008}  as explicitly shown in fig.\ref{fig:xsplus}. In the same figure the NLO and LO evolution predictions (obtained starting from the same low resolution scale) are also shown for comparison in order to demonstrate the large NNLO effect \cite{notestrange}. Of course if one assumes zero asymmetry at $Q_0^2$ the LO and NLO predictions would vanish identically. One can look at the table \ref{table1}, to compare the present results with the first  moments of the strange distributions evaluated within a lattice QCD approach \cite{LATasymmetry2009} which, however,  is subjected to large systematic errors.
It is evident from the results of fig.\ref{fig:xsplus} and table \ref{table1} that the asymmetry is not well constrained at NLO, a conclusion which is rather contradictory with recent NLO approaches  which attempt to extract the strange asymmetry from NLO analysis (e.g. ref.\cite{s_minusLO}). A NNLO investigation is mandatory and it is one of the privileges of the present approach the investigation of the strange asymmetry from non-perturbative sources and NNLO accuracy. The conclusion is rather simple: the total strange distribution $x (s(x) + \bar s(x))$ is well reproduced (cfr. fig.\ref{fig:xsplus}) and the asymmetry is characterized by negative moments as shown in table \ref{table1}. The reason has to be ascribed to the meson-cloud modeling of the strange sector and it is emphasized by the NNLO evolution. In fact the present conclusions have large common features with other recent results on the strange distribution effects within meson cloud in non-perturbative models.  In ref.\cite{SongZhangMa2010}, for example, the chiral quark model is shown to explain the Gottfried sum rule, but it predicts corrections to the NuTeV anomaly which are in opposite direction (as in the present investigation), a conclusion already established in ref.\cite{DingMa2006} where the corrections, calculated within a meson cloud model, were found very small and opposite in sign.
Signal {\sl et al.}  in ref.\cite{Signal_etal2010}) stress that even the introduction of additional components due to $K^*$ states cannot invert the conclusions.

\begin{table*}[btp]
%\addtolength{\tabcolsep}{20pt}
\addtolength{\extrarowheight}{5.0pt}
\caption{Moments of the strange distributions calculated within the scenarios {\bf A} and {\bf B} and evolved at NNLO ($Q^2 = 4\,{\rm GeV^2}$) are compared with the results of the lattice calculation of ref.\cite{LATasymmetry2009} (the authors warn that the their calculation is subjected to large systematic errors). The results of the NLO evolution are shown in parenthesis $[...]$.  
}
\begin{ruledtabular}
\begin{tabular}{lllll}
\\[-2.0em]
 & $\langle x (s(x) + \bar s(x)) \rangle$ & $\langle x (s(x) - \bar s(x)) \rangle $  & $\langle x^2 (s(x) - \bar s(x) )\rangle $  \\ 
 {\bf A}: no strange sea at $Q_0^2$ &$0.046\, [0.024]$ &$-0.0015\, [0]$ &$-0.00018\, [0]$  \\
 {\bf B}: strange sea at $Q_0^2$&  $0.051\,[0.036]$ &$-0.0039\, [-0.0033]$ &$-0.0014 \,\;\;[-0.0017]$  \\ 
\hline
from ref.\cite{LATasymmetry2009} &$0.027 \pm 0.006$ & consistent with zero & consistent with zero 
\end{tabular}
\end{ruledtabular}
\label{table1}
\end{table*}

%%%%%%%%%%%%%%%%%%%%%%%      table 1

%===============================================================================
\begin{figure}[tbp]
%\vspace{-0.25truecm}
%\centering\includegraphics[width=\columnwidth,clip=true,angle=0]{NNLO-NLO-LO-Hermes-strange.eps}
\centering\includegraphics[width=\columnwidth,clip=true,angle=0]{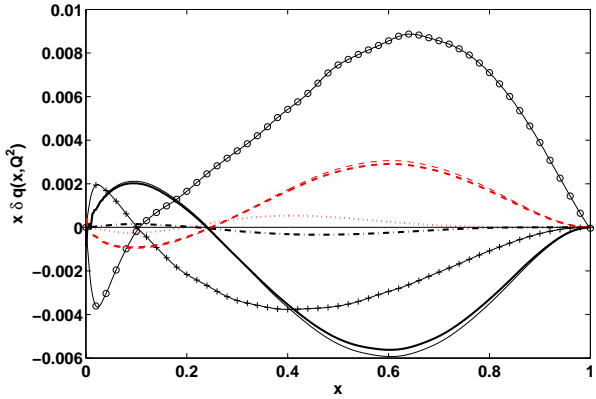}
%\centering\includegraphics[width=\columnwidth,clip=true,angle=0]{Rminus100000.eps}
\caption{Isospin symmetry violations from radiative QED effects (from eqs.(\ref{gammaintegrated}) at $Q^2 = 10\,{\rm GeV^2}$) and mass effects (from the model (\ref{LFmodel})  at $Q_0^2$). $x \delta u_V(x,Q^2)$ (continuous lines, the tiny line does not include strange sea at the static point $Q_0^2 = 0.149\,{\rm GeV^2}$) and $x \delta d_V(x,Q^2)$ (dashed lines, the tiny line does not include strange sea at the static point $Q_0^2$).
$x \delta \bar u$ and $x \delta \bar d$ are represented by the dot-dashed and dotted lines respectively, they are calculated including strange sea at the static point. The effects due to the $u - d$ mass difference ($m_d-m_u = 4\,{\rm MeV}$ according to ref.\cite{LoTho2003}), are shown by line-circles ($x \delta d_V(x,Q_0^2$)) and by line-plus  ($x \delta u_V(x,Q_0^2)$).
}
\label{fig:xdeltaq}
\vspace{-1.0em}
\end{figure}
%===============================================================================

%%%%%%%%%%%%%%%%%%%%%%%%   table 2

\begin{table*}[btp]
%\addtolength{\tabcolsep}{20pt}
\addtolength{\extrarowheight}{5.0pt}
\caption{Corrections to $\Delta s_W^2$ calculated according to (\ref{PWexpansion}) and (\ref{functional}) within the LF model described in the text. Scenarios {\bf A} and {\bf B} are presented for the meson cloud. The results refer to the experimental scale $Q^2 = 10\,{\rm GeV^2}$ and the evolutions are performed at NNLO. Comparison with the recent reassessment of ref.\cite{Bentz-etal2010} is proposed.}
\begin{ruledtabular}
\begin{tabular}{lllllllll}
\\[-2.0em]
contribution to  &  from   & from   & from    & from   &  from   &   from    &       &  \\        
$\Delta R^{CSV}$ & $\left. x \delta u_V\right|_{mass}$  & $\left. x \delta d_V\right|_{mass}$ & $\left. x \delta u_V\right|_{QED}$  & $\left. x \delta d_V\right|_{QED}$ & $\left. x \delta \bar u\right|_{QED}$ & $\left. x \delta \bar d \right|_{QED}$ & Total $\Delta R^{CSV}$& \\ 
\hline
{\bf A}: Êno strange sea at $Q_0^2$& $-0.0003$&$-0.0005$ &$-0.0006$& $-0.0003$ &$-0.000004$ &$-0.000008$& $-0.0017$  &             \\
%\hline
{\bf A}: no strange sea at $Q_0^2$ & $-0.0005_{PW}$ & $-0.0013_{PW}$ & $-0.0016_{PW}$ & $-0.0008_{PW}$  & $-0.0001_{PW}$ &  $-0.0001_{PW}$  & $-0.0044_{PW}$ &		     \\
%\hline 
{\bf B}: strange sea at $Q_0^2$& $-0.0003$&$-0.0006$ &$-0.0006$& $-0.0003$ &$-0.000004$ &$-0.000008$ & $-0.0018$  & 
\\
%\hline 
{\bf B}: strange sea at $Q_0^2$ & $ -0.0005_{PW}$ & $-0.0013_{PW}$ & $-0.0015_{PW}$ & $-0.0008_{PW}$  & $-0.0001_{PW}$  & $-0.0001_{PW}$  & $-0.0043_{PW}$ &	 \\
%\hline
\end{tabular}
\begin{tabular}{lcccccccccc}
present work &&& $\underbrace{-0.0009 \pm 10\%}_{mass}$ &   &  & $\underbrace{-0.0009}_{QED}$  &    & & $ \underbrace{-0.0018 \pm 0.0001}_{{\rm total}\; CSV}$ & \\ 
present work &&& $\underbrace{-0.0018_{PW}}_{mass}$ &   &  & $\underbrace{-0.0025_{PW}}_{QED}$  &    & & $ \underbrace{-0.0043_{PW}}_{{\rm total}\; CSV}$ & \\ 
from ref.\cite{Bentz-etal2010}   & &  & $\underbrace{-0.0015 \pm 20\%}_{mass}$ &   &  & $\underbrace{-0.0011 \pm 100\% }_{QED}$  &    & & $ \underbrace{-0.0026 \pm 0.0011}_{{\rm total}\; CSV}$ & \\ 
from  ref.\cite{Bentz-etal2010}   &  & &$\underbrace{-0.0020_{PW}}_{mass}$   &   & &  $\underbrace{-0.0020_{PW}}_{QED}$     &       &       & $\underbrace{-0.0040_{PW}}_{{\rm total}\; CSV}$ &
\end{tabular}
\begin{tabular}{lllllllll}
  $\Delta R^i$ &  & $[\Delta R^{N,Z} - \Delta R^{N,Z}_{\rm NuTeV}]$ &  $\Delta R^{CSV}$ & $\Delta R^{\rm strange}$  & & $\Delta s_W^2 = \sum_i \Delta R^i $ & \\
\hline
{\bf A}: no strange  sea at $Q_0^2$  &  & $-0.0013$  & $-0.0017$ & $+0.0011$ &     &  \\ 
{\bf A}: no strange  sea at $Q_0^2$  &  & $-0.0035_{PW}$& $-0.0044_{PW}$ & $+0.0020_{PW}$ &     &  \\ 
{\bf B}: strange sea at $Q_0^2$       &  &$-0.0013$  & $-0.0018$ &  $+0.0028$ &     &   \\ 
{\bf B}: strange sea at $Q_0^2$       &  & $-0.0035_{PW}$ & $-0.0043_{PW}$ &  $+0.0047_{PW}$ &     &   \\ 
\hline
present work &   & $-0.0013 \pm 0.0001$ &  ${-0.0018 \pm 0.0001}$ & $\phantom{+}0.0 \pm 0.0011$&      & \fbox{$-0.0031 \pm 0.0011$}     \\
present work &   & ${-0.0035 \pm 0.0003}_{PW}$ &  ${-0.0043 \pm 0.0001}_{PW}$ & $\phantom{+}{0.0 \pm 0.0020}_{PW}$&      & \fbox{${-0.0078 \pm 0.0020}_{PW}$} 
\\ from ref.\cite{Bentz-etal2010}         &   & $-0.0019 \pm 0.0006$ &  ${-0.0026 \pm 0.0011}$ & $\phantom{+}0.0 \pm 0.0018$&      & $-0.0045 \pm 0.0022$     \\ 
from  ref.\cite{Bentz-etal2010}        &   &$ {-0.0025_{PW}}$ &   ${-0.0040}_{PW}$  & ${\phantom{+}0.0}_{PW}$  & &$-0.0065_{PW}$         \\ 
\end{tabular}
\end{ruledtabular}
\label{table2}
\end{table*}

%===============================================================================
\begin{figure}[tbp]
\centering\includegraphics[width=\columnwidth,clip=true,angle=0]{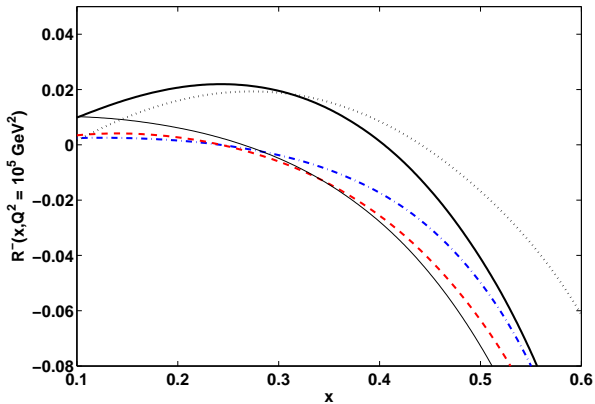}
\caption{$R^-(x,Q^2 = 10^5\,{\rm GeV^2})$ from eq.(\ref{eq:Rminus}) as function of $x$. The dot-dashed curve shows the contribution due to the QED splitting only;  the dashed curve includes the additional contribution from quark mass CSV terms; solid curve: complete results adding all terms including strange quark asymmetry at NNLO (scenario {\bf B}).  The same results at NLO are shown by the dotted line.
Predictions at NNLO, obtained neglecting strange components at the initial scales  (scenario {\bf A}) are summarized by the solid tiny line. }
\label{fig:Rminus}
\vspace{-1.0em}
\end{figure}
%===============================================================================%%%%%

The additional contribution to the valence isospin asymmetries stemming
from radiative QED effects was presented in ref.\cite{GJDR2005}.
Following the spirit of this publication one can  evaluate $\delta q_V$
and $\delta\bar{q}$ utilizing the QED ${\cal{O}}(\alpha)$
evolution equations for $\delta q_V(x,Q^2)$ and $\delta\bar{q}(x,Q^2)$
induced by the photon radiation off quarks and anti-quarks.
To leading order in $\alpha_{EM}$  one has
\begin{eqnarray}
\frac{d}{d\ln Q^2}\, \delta u_V(x,Q^2) & = & + \frac{\alpha_{EM}}{2\pi}\int_x^1
 \frac{dy}{y}\,P\left(\frac{x}{y}\right) u_V(y,Q^2)\nonumber\\
\frac{d}{d\ln Q^2}\, \delta d_V(x,Q^2) & = & -\frac{\alpha_{EM}}{2\pi}\int_x^1
 \frac{dy}{y}\,P\left(\frac{x}{y}\right) d_V(y,Q^2)\,,\nonumber \\
 \label{gammaevolution}
\end{eqnarray}
with $P(z) = (e_u^2-e_d^2)P_{qq}^{\gamma}(z) = (e_u^2-e_d^2)
\left(\frac{1+z^2}{1-z}\right)_+$.
Similar evolution equations hold for the isospin asymmetries of 
sea quarks $\delta\bar{u}(x,Q^2)$ and $\delta\bar{d}(x,Q^2)$. 
The integration is performed in the following way:
\begin{eqnarray}
\delta u_V(x,Q^2) & = & + \frac{\alpha_{EM}}{2\pi} 
      \int_{m_q^2}^{Q^2}d\ln q^2 
       \int_x^1\frac{dy}{y}\,\, 
         P\left(\frac{x}{y}\right) u_V(y,\, q^2)\nonumber\\
\delta d_V(x,Q^2) & = & -\frac{\alpha_{EM}}{2\pi}
      \int_{m_q^2}^{Q^2}d\ln q^2 
       \int_x^1\frac{dy}{y}\,\, 
         P\left(\frac{x}{y}\right) d_V(y,\, q^2) \nonumber \\
         \label{gammaintegrated}
\end{eqnarray}
and similarly for $\delta\bar{u}$ and $\delta\bar{d}$. $q_V(x,Q^2)$
and $\bar{q}(x,Q^2)$ are the distributions  of the light-cone quark model previously discussed, and 
the current quark mass $m_q$  is  conservatively chosen $m_q=10$ MeV \cite{GJDR2005}. 
The parton distributions at  $\left. Q^2< Q_0^2\right|_{\rm LO}$ in (\ref{gammaintegrated}) 
($\left. Q_0^2\right|_{\rm LO}$ is the input scale \cite{Traini2011}) 
are taken to equal their values at  $\left. Q_0^2\right|_{\rm LO}$, 
$\stackrel{(-)}{q}\!\!(y,\, Q^2\leq \left. Q_0^2\right|_{\rm LO}) =
\stackrel{(-)}{q}\!\!(y,\, \left. Q_0^2\right|_{\rm LO})$, i.e.\ are assumed \cite{GJDR2005} `frozen' \cite{note2}.

In fig.\ref{fig:xdeltaq} the results for all the (independent) CSV contributions are summarized. One can appreciate  the relative effects of the $u-d$ mass difference as evaluated within LF model (\ref{LFmodel}) and  the QED contributions (\ref{gammaintegrated}) to both valence (eq.(\ref{CSpartonicV})) and sea (eq.(\ref{CSpartonicSea})) distributions.
One has $\left. \langle x \delta u_V\right|_{mass} \rangle = -0.0018$, $\left. \langle x \delta d_V\right|_{mass} \rangle = 0.0044$ at $Q_0^2$, \cite{staticscale}.

\vspace{0.1truecm}

\noindent{\bf 5. results, further discussion and perspectives}
%\vspace{0.1truecm}

It is important to recall that in the NuTeV analysis the measured quantities (NC to CC ratios for the neutrinos and antineutrinos of eq.(\ref{PWratio})),  are related to the extracted values of the weak mixing angle through a Monte Carlo analysis. For a given effect, the $PW$ ratio ($\left. \Delta s^2_W\right|_{PW} \equiv \Delta R_{PW}^i$, cfr. eqs.(\ref{PWexpansion}),  will only give a qualitative estimate of the actual value of the corrections.  Quantitative corrections are obtained by using the functionals  
\begin{equation}
\Delta s^2_W = \int^1_0 F[s^2_W, \delta \!\!\stackrel{(-)}{q}\!\!; x]\; x \delta \!\!\stackrel{(-)}{q}\!\!(x,Q^2=10\,{\rm GeV^2})\, dx \label{functional}
\end{equation}
provided in ref.\cite{NuTeVdata} by the NuTeV collaboration: the resulting corrections will be denoted $\Delta s^2_W \equiv \Delta R^i$ and shown in tables (\ref{table2}) where also the corresponding ${PW}$- estimates ($\Delta R_{PW}^i$) are shown and labeled with the subscript $..._{PW}$. A comparison of the present results with the reassessment of the NuTeV corrections proposed by Bentz {\sl et al.}\cite{Bentz-etal2010} is presented in the same table \ref{table2}.

\vspace{0.05truecm}

\noindent - {\sl $\Delta R^{N,Z}$ correction}: it is related to the non isoscalar  nature of the target and it has been (in principle) taken into account by the NuTeV collaboration. Its value is phenomenologically rather well identified and large (the value $\Delta R^{N,Z}_{\rm NuTeV} = 0.008 \pm 2\%$ is sensible\cite{CSV2010} using the recent PDF fits). Two comments: i) the present study is not using PDF fits, but models of them, the correction has to be evaluated consistently within the model and a shift of  $-0.0013 \pm 0.0001$ is obtained and shown in table \ref{table2}; ii) $\Delta R^{N,Z}$ is influenced by nuclear corrections which take into account the effects of the neutron excess trough EMC effect \cite{Cloet-etal2009}. The assumption of incoherent scattering as it emerges from eq.(\ref{incoherent}) is no longer valid. A correction of $-0.0019 \pm 0.0006$ has been indicated \cite{Bentz-etal2010}. The present approach can be used to study also such a correction \cite{EMC-LF},  however, to be conservative, table \ref{table2} includes the shift due to the incoherent parton model analysis only.
\vspace{0.05truecm}

\noindent - {\sl $\Delta R^{CSV}$ correction}: the evaluation of the CSV components due to $m_d-m_u$ mass dfference and QED effects is, for the first time, developed at NNLO and the results detailed in the upper part of table \ref{table2}. The comparison with the reassessment \cite{Bentz-etal2010} shows a numerically consistent agreement with recent PDF fits. In particular the result of the present paper $\Delta R^{QED}_{PW} = - 0.0025$ with the estimate of the MSRT group \cite{MRST2005} who explicitly included the QED splitting effects in PDF evolution and found $\Delta R^{QED}_{PW} = - 0.0021$ at $Q^2 = 20\,{\rm GeV^2}$. 
$\Delta R^{CSV}$ is probably the better established correction.

\noindent - {\sl $\Delta R^{\rm strange}$ correction}: it remains the most critical correction because connected to the strange asymmetry which is definitely poorly known. The comparison of the present results with lattice calculations as made in table \ref{table1} shows a rather consistent agreement for the strange sea $\langle x(s+\bar s \rangle_{Q^2 = 4\,{\rm GeV^2}}$, (within the large systematic errors which affect the lattice evaluation), and a strong NNLO correction. The actual NNLO values strongly depend on the (also) poorly known distribution of the strange sea at the  non-perturbative scale showing that one cannot remain within a NLO analysis to extract realistic distributions. It is physically sensible also in the present investigation to follow the proposal of ref.\cite{Bentz-etal2010} putting to $0.0$ the asymmetry contribution and enlarging the systematic error. A solution that takes into account the analysis of recent cloud-model studies like those ones of refs.\cite{SongZhangMa2010,DingMa2006,Signal_etal2010}, but which, at the same time, urges to new theoretical and experimental investigations.

\vspace{0.05truecm}

\noindent - {\sl $\Delta s_W^2$}: the global corrections proposed at the end of the present NNLO analysis
are summarized in the boxes of table \ref{table2} and one obtains:
\begin{equation}
\sin^2 \theta_W = 0.2246 \pm 0.0013\,(stat)  \pm 0.0020\,(syst)\;.
\end{equation}
 The discrepancy is reduced within less than $1\, \sigma$
and, in view of the nuclear corrections, the NuTeV anomaly is solved on a completely theoretical basis.

\vspace{0.05truecm}

\noindent - {$R^-(x)$}: quite recently the possibility of measuring CSV by comparing neutrino or antineutrino production through charged-current reactions induced by electrons or positrons at a possible electron collider at the LHC (LHeC \cite{http_uk}) has been examined \cite{Hobbs_etal2011}.  The magnitude of the CSV effects that might be expected at such facility is likely to be of several percent, substantially larger than the typical CSV effects expected for partonic reactions. In particular in ref.\cite{Hobbs_etal2011} the observable
\begin{eqnarray}
R^-(x) & \equiv & {2 [F_2^{W^- D}(x) - F_2^{W^+ D}(x)] \over F_2^{W^- p}(x) - F_2^{W^+ p}(x)} = \nonumber \\
& = &  {x[-2 s^-(x) + \delta u^-(x) - \delta d^-(x)] \over x[u^+(x) + d^+(x) + s^+(x)]}\,,
\label{eq:Rminus}
\end{eqnarray}
is defined (in eq.(\ref{eq:Rminus}) small charm contributions have been neglected). The quantity (\ref{eq:Rminus}) is given by the difference in the $F_2$ structure functions per nucleon for electron-deuteron and positron-deuteron CC reactions, divided by the average $F_2$ structure function for CC reactions on protons initiated by electrons and by positrons. $R^-$ is sensitive to CSV effects as it borns out from the figure \ref{fig:Rminus} where the effects of the various terms of eq.(\ref{eq:Rminus}) are shown for an 
hypothetical experiment proposed in ref.\cite{Hobbs_etal2011} at the mentioned collider. The sensitivity  of $R^-$ to the strange asymmetry and the evolution approximation is emphasized. .The electron/positron beams (of energies in the range $50 - 100\,{\rm GeV}$) interact with deuteron beams at LHC  energies of several TeV producing the most promising observables with which to search for partonic charge symmetry violating effects.

%%%%%%%%%%%%%%%%%%%%%%%%%%%%%%%%%%%%%%%%%%%%%%%%%%%

\vspace{0.1truecm}

\noindent{\bf  acknowledgements}

%\begin{acknowledgements}

I thank M. Garcon, head of  the Service de Physique Nucl\'eaire, and B. Shagai for kind hospitality at SPhN - CEA, Saclay were I found the needed concentration to complete my NNLO evolution code. Useful information from B. Pasquini are also gratefully  acknowledged.

M.T. is member of the Interdisciplinary Laboratory for Computational Science (LISC), a joint venture of Trento University and FBK. 

The present research is partially funded by the Provincia Autonoma di Trento and by INFN, through the AuroraScience project.  

%\end{acknowledgements}


\begin{thebibliography}{50}
\bibitem{staticscale} The isospin violation terms due to $m_d-m_u$ mass difference, are evaluated at the model scale $Q_0^2$. Their contribution to the NuTeV results is basically independent from the evolution scale because it is a ratio of non singlet distributions  (cfr. eq. (\ref{PWexpansion})) evolving in the same way.
\end{thebibliography}
\end{document}